\documentstyle[manuscript]{aastex}
\def \CQ             {{1997~CQ$_{29}$}}
\def \CQF           {{2000~CQ$_{114}$}}
\def \WW             {{1998~WW$_{31}$}}

\def \CF             {{2000~CF$_{105}$}}

\def \QC             {{2001~QC$_{298}$}}
\def \QG             {{2001~QG$_{298}$}}

\def \RZ             {{1999~RZ$_{253}$}}

\def \CM             {{(80806)~2000~CM$_{105}$}}

\def \CS             {{(79360)~1997~CS$_{29}$}}

\def \OJ             {{1999~OJ$_{4}$}}
\def \YW             {{(82075)~2000~YW$_{134}$}}
\def \OJS             {{2000~OJ$_{67}$}}

\def \TL             {{(48639)~1995~TL$_{8}$}}

\def \ph          {\phantom}

\def \ref            {\noindent\hangindent0.5in\hangafter=1}

\begin{document}

\title  {Detection of Six Transneptunian Binaries with NICMOS:  A High Fraction of Binaries in the Cold Classical Disk}

\author{Denise C.~Stephens}
\affil{Johns Hopkins University, Dept.~Physics and Astronomy, Baltimore, MD 21218}
\author{Keith S.~Noll} 
\affil{Space Telescope Science Institute, 3700 San Martin Dr., Baltimore, MD 21218}
\email{stephens@pha.jhu.edu, noll@stsci.edu}

\newpage

\begin{abstract}  

We have analyzed a homogeneous set of observations of eighty-one transneptunian objects obtained with the NIC2 camera on the Hubble Space Telescope with the goal of identifying partially resolved binaries.  Using PSF-fitting we have identified six likely binaries in addition to the three new binaries already found in this data set.  We find that 11$\pm {5\atop 2}$\% of transneptunian objects are binaries at separation and brightness limits of the NIC2 camera.  The identification of these new binaries significantly increases the known lower limit to the binary fraction among transneptunian objects.  The origin of such a high fraction of binaries remains to be determined.  Most interestingly, detectable binaries appear to be about four times more common among the cold classical disk than in the dynamically excited populations.

\end{abstract}

\keywords{Kuiper Belt, Oort cloud}

\newpage
\section{Introduction}

Populations of bodies interacting under the influence of mutual gravitation can produce a variety of interesting dynamical outcomes.  Collisions are the most dramatic example of such an interaction.  More common, however, are close encounters in which the mutual gravitation of a pair of objects becomes temporarily significant.  Depending on the space density of the population of bodies, more complex interactions can occur involving three or more objects leading to less predictable and more complex dynamical outcomes.  One possible outcome of these interactions is the production of gravitationally bound systems of two or more objects.

The discovery of Pluto's moon Charon (Christy and Harrington 1978) was, in retrospect, the first discovery of what we now know to be a substantial population of transneptunian binaries (TNB).  Roughly a decade after the discovery of the Kuiper Belt, Veillet et al.~(2002) found the next binary, a companion to \WW , among the now-expanded population of transneptunian objects.  This has been followed by a rapid pace of discovery with confirmed TNBs now numbering fifteen (Noll 2003,Kern and Elliot 2005, Brown et al.~2005).

In addition to the obvious practical application of observing binary orbits for mass determination, an important first-order question is how many binaries exist.  The answer is subject to many poorly-quantifiable observational biases.  It is clear that a minimum of several percent of transneptunian objects, in the aggregate, are multiple (Noll 2003).  How many more might be binary remains to be determined by higher angular resolution observations or more careful data analysis of existing data.  In this work, we describe a systematic analysis of a large, homogeneous, high-resolution data set obtained with the Hubble Space Telescope that is aimed at revealing partially-resolved binaries that have not been previously identified.

\section{Observations}

Most of the data were obtained by the NIC2 camera between August 6, 2002 and June 25, 2003.   The pixel scale of the NIC2 camera is 0.0759 arcsec/pixel in $x$ and 0.0754 arcsec per pixel in $y$.  Observations were made in two filters, the F110W ($\sim$J band) and the F160W ($\sim$H band) in the standard multiaccum imaging mode.  The observing sequence consisted of a 256 second exposure in J, a 512 second exposure in H, a dither of 5.5 pixels in both $x$ and $y$ directions, followed by a 512 second exposure in H and finally a 256 second exposure in J.  The apparent motion of the TNO caused both by proper motion and parallax was tracked by the spacecraft so that background stars and galaxies are trailed.  Unambiguous identification of the TNO was possible both from its lack of motion and from the consistently good pointing that typically put the target within an arcsecond of the expected location on the detector.  A total of seventy-seven unique objects not previously known to be binary were successfully observed during this year-long campaign.  From visual inspection of the data we found three of these seventy-seven to be binary objects, \QC , \RZ , and \CQF\ (Noll et al.~2002a, 2003; Stephens et al.~2004). 

We also analyzed four additional objects observed with the NIC1 and NIC2 cameras between June 16, 1998 and September 10, 1998 as part of an earlier program described by Noll et al.~(2000).  For the sake of maintaining a homogeneous sample, we consider here the PSF-fitting to the NIC2 data obtained in the F110W and F160W filters, identical to the filters used in the larger sample.  Adding these four to our sample brings the total to eighty-one objects which were not known to be binary prior to the work described in this paper.  This sample of eighty-one forms the largest, most homogenous, fully-sampled data set on which the PSF-fitting technique can be used.

For the sake of completeness, we note that our NICMOS program included three additional orbits devoted to previously known binaries;  these are not counted in the sample of eighty-one objects detailed above.  One binary, \CF , previously found with WFPC2 (Noll et al.~2002b) was observed in two separate orbits with NICMOS.  One of the three new binaries found in our NICMOS data, \QC , was scheduled for a repeat observation by us after its discovery.  The circumstances for all of the objects that are confirmed, visually resolved binaries, whether known to be binary before the NICMOS observations or not, are summarized in Table~1.  All of the binaries except \QC\ had separations of the two components at the time of discovery of more than two pixels, consistent with what we would expect for a search based on visual inspection.  

\section{Analysis}
\subsection{PSF Fitting}

A great advantage of observations with HST is the stability of the point spread function (PSF) during the course of the observation.  This stability means that it is possible to identify binary systems that are not resolved according to the usual Rayleigh criterion.  In the filter bandpasses we used, the diffraction limited central peak of the Airy distribution for a point source has a FWHM of 115 milliarcsec (F110W) and 168 milliarcsec (F160W), equivalent to 1.5 and 2.2 pixels.  As Table~1 demonstrates, no target with separation of less than 1.7 pixels was identified as a double in our initial pass through the data, consistent with what would be expected.   With PSF-fitting, however, we can identify binaries by the changes they induce to the combined PSF at separations  considerably smaller following the method described below.  

Model PSFs were generated using the TinyTim software (Krist and Hook 2003).   For each pixel in the NICMOS image we considered a grid of 100 model PSFs shifted by 0.1 pixel (0.0075 arcsec) in $x$ and $y$.  The program proceeds by successively testing pairs of possible positions for the primary and the secondary.  The range of pixels that were considered was constrained to be those pixels with signal above the background, typically a square with 4 or 5 pixels on a side.  At each possible combination of positions for the primary and secondary, we iteratively scaled each PSF while preserving the overall flux and determined the residuals.  This process was repeated until we found the combination that minimized the residual for each position pair.  The position pairing that gives the lowest minimized residual is considered to be the best binary solution.  It remains to be determined, however, whether this best binary solution is to be preferred to a simpler, single component solution which we calculate by a similar PSF fitting process.  

An advantage of considering each of the four images separately is that we can rule out false detections due to tracking errors or transient orbital events.  High jitter events on HST in 3 gyro control are rarely of an amplitude greater than 0.010 arcsec and have short durations.  Thus, even if one image were to be affected, the abnormality of the PSF would be small and confined to a single exposure.  Similarly, PSF-blur caused by an incorrect track rate would be identifiable as a slow drift in position between exposures.  There is no evidence for any such effects in our data.  We note that existing PSF fitting routines like DAOPHOT generally require much higher signal than is available in our data.

\subsection {Identification of Candidate Binaries}

To determine whether the results of the binary fitting procedure indicated the presence of an unresolved binary, we established five criteria intended to distinguish between good candidates and chance numerical results.  In the NICMOS sample we have four images, two each in the F110W and F160W filters.  All of the observations were dithered with images taken at two different positions on the chip.  For the bulk of the data, the first F110W (J) and F160W (H) image pair is followed by a 5.5 pixel offset in $x$ and $y$ and exposures in F160W and F110W.  For the data obtained in 1998, the images were dithered by 5 pixels in $x$.  We fit each of the four NICMOS images individually using the binary PSF program, and then compare the positions, separations, and relative fluxes of each binary solution.  We applied the following criteria:

\begin{itemize}

\item The separation of the primary and secondary must agree to 0.01 arcsec or better in all images.

\item The position angle of the secondary relative to the primary must agree to within 10 degrees for all images.

\item The fraction of flux contributed individually by the primary and secondary to the total flux of the system cannot vary by more than 10\% .

\item The coordinates found for the primary and secondary in the first two images must be shifted from the second two images by $\approx$5.5 pixels.  (In two cases, as noted in Table~2, one of the two dithered positions falls on bad pixels and cannot be used)

\item The $\chi ^2$ of the best binary fit must be better by more than 3$\sigma$ than the $\chi ^2$ of the best single object fit.   

\end{itemize}

Our analysis identified six of the seventy-eight  apparently single objects that have binary solutions meeting all of the criteria.  These are listed in Table~2.  The separations we find for these objects range from 0.2 to 1.2 pixels with all but one at 0.8 pixels or more.  We note that two of the identifications rely on just two exposures at a single pointing because one of the two dithered pointings put the TNO image on a photometrically defective pixel, rendering that J and H pair unusable for PSF fitting. 

Figure 1 shows one of our candidate binaries, \CM , compared with the best-fitting binary model.  Below these panels are the residuals after subtracting the best binary and single model.  In the central 16 pixels the binary-fit residuals are consistent with the background noise; the residuals from the single-object fit are a factor of two per pixel greater in the same region.   The $\Delta\chi ^2$ for the binary solution compared to the best single object solution indicates a 4$\sigma$ or greater confidence level for the binary solution over the single solution for each of the four exposures.  (We have assumed that both single and binary solutions consist of six degrees of freedom, the $x$ and $y$ separation of both the primary and the secondary and the scaling factor $z$ for each.  Single fits are considered to be cases where $z = 0$ for the secondary.)

For the sake of completeness, we have listed in Table~3, all of the objects successfully observed with NICMOS which do not have evidence for a binary companion.  It is important to note that the inclusion of an object here does {\em not} mean that it is not a binary, but only that at the epoch of the observation there was no evidence of a binary of a magnitude and separation sufficient for detection according to the criteria we have established.  It could well turn out that future observations reveal some of these objects to have companions that were too small, too close, or that were missed because of unlucky placement of cosmic rays or bad pixels.  

\subsection{Testing}

As a first test, we computed the separations of the resolved binaries listed in Table~1 using the PSF fitting code and compared the results to separations obtained by image centroiding.  In all such cases, we find excellent agreement between the results regardless of method, though the PSF fitting routinely results in smaller uncertainties.  All of the binary separations listed in Table~1 have been recalculated using our PSF fitting program and supersede previously published values where these differ.  Both NICMOS observations of  \CF\ and the second epoch observation of \QC\ have not previously been reported. 

It is worth noting that, although the first epoch observation of \QC\ had a separation of nearly two pixels (Noll et al.~2002a), the second epoch observation had a separation of only slightly more than one pixel.  This second epoch observation originally led us to doubt the reality of our earlier claim of detection for this system.  However, using the binary PSF fitting program we are easily able to identify this object as a binary at both epochs and find separations with uncertainties on the order of a tenth of a pixel.  In Table~1 we report separations of this pair based on PSF fitting.  For the October 2002 epoch, the PSF fitting results in an order of magnitude reduction in the uncertainty going from 0.17$\pm$0.08 arcsec as originally reported by us (Noll et al.~2002a) to 0.130$\pm$0.007 arcsec.  In April 2003 we find a separation of 0.097$\pm$0.009 arcsec.

 The second test method we used requires the creation of artificial binaries over a range of radial separations, position angles, and relative intensities to determine detection probabilities.  We constructed test images by adding scaled PSFs with known relative positions and flux to real NICMOS backgrounds.  These images were then used as inputs to the binary PSF-fitting program.  
 
For each primary magnitude a series of test images were created with the secondary differing from the primary in half-magnitude steps from a $\Delta$mag=0 (two identical components) to $\Delta$mag=3.  For each primary magnitude-$\Delta$mag pair, five separate test images were produced.  Each of these five had a different background noise pattern and a different sub-pixel centering of the primary.  In all, we tested more than 7000 constructed images to establish approximate detection limits for this experiment.

The same criteria were used to determine whether or not the search program had reliably found the embedded binary in the artificial binaries as in the real data.  A subset of the results are shown in Figure~2.  In this figure we plot sets of points for four different primary magnitudes.  The separation plotted at each half-magnitude step in $\Delta$magnitudes is the minimum distance at which half or more of the test images resulted in a positive detection of the secondary.   The results plotted in Fig.~2 show some noisiness due to the relatively low number of samples per point.  However, the test data are sufficient to establish the reliability of the detections made in real NICMOS data.

Two features are especially notable from the results of these tests.  First, from a comparison of Fig.~2 and Table~2 it is apparent that all of the objects are within the regions of detectability for an object of their brightness.  The two faintest objects, \CM\ and \OJ\ are closest to the boundary and so it is particularly interesting to note that we have recently acquired new images of \OJ\ with the ACS HRC that resolve the binary pair (Fig.~3).  Interestingly, the very small separation of \TL\ is detectable because of the brightness of the object pair.  

A second important feature that can be seen in Fig.~2 is that the minimum separation for detection of binary companions steadily increases as the brightness of the primary decreases.  This arises because the companions are approaching the detection limit for any object with the integration times used in these observations, approximately 23.5 mag in the F160W filter.  The faintest objects with a detected companion in our sample have an F160W mag of 21.2.  The median magnitude for our sample is 21.15 mags which means that none of the objects in the fainter half of our sample (which extends to F160W = 22.5) has a detected companion.  This lack of detection may be due, at least in part, to the limited sensitivity of the NICMOS observations.  It is important to note that there are {\em no} biases of brightness by dynamical class.  The median magnitude for all the classical, resonant, and scattered objects in our data set are 21.2, 20.9, and 21.1 mags respectively.

It is important to note that the $\Delta$mag shown in Table~2 is intended to indicate the reliability of the detection and not the actual difference in magnitudes of the components.  This value is subject to systematic uncertainties in the fitting process and it is therefore difficult to quantify the uncertainty of this value.  In the case of \OJ\ the ACS HRC data show a difference between the components that is smaller than suggested by Table~2.
 

\section {Discussion}

One of the most basic questions about binaries in the Kuiper Belt that has yet to be answered is how many exist.  The fraction of objects in each of the dynamical classes that are binary systems is a function of the conditions that led to the formation of binaries at an early epoch in the solar system and their subsequent disruption over 4.5 billion years.  Because of the faintness and small angular separation of transneptunian binaries, there are many potential observational biases limiting our ability to answer this question.  

The NICMOS target list was constructed based on the availability of TNO orbits sufficiently well-determined that the uncertainty in position during the observing period was less than 5 arcseconds.  This is a random criterion with regard to the binarity of a particular object.  We thus consider the NICMOS target list to be unbiased for binaries.  The only exception is for visits that were specifically added after a binary had been identified. Two objects, \CQ\ and \CF\ were discovered as binaries with WFPC2 (Noll et al.~2002b) and were known to be binary at the time the NICMOS target list was constructed.  \CQ\ was observed once with NIC2 and \CF\ was observed twice.  In addition to the two WFPC2-identified systems, \QC\ was identified as a binary from NICMOS data (Noll et al.~2002a) and a second visit was added to the NICMOS target list following this discovery.  One object, 2000 PV$_{29}$, was inadvertently observed twice, once as 2000 PV$_{29}$ and once as minor planet 45802.  Thus, of the eighty-six successful NICMOS observations, only eighty-one distinct TNOs can be considered unbiased with regard to binaries.  

In this population of eighty-one TNOs we have now identified a total of nine binary systems, six of which we identify here for the first time.  Considering, for the moment, the entire ensemble, we derive an ``average'' binary fraction of 11$\pm {5\atop 2}$\%.  This is almost a factor of three higher than our earlier estimate from this data set, prior to the detections reported here (Noll 2003).  As discussed above and as the separations in Tables 1 and 2 demonstrate, this is entirely due to sampling a larger part of the available search space near each object.  

As our tests reveal, the detection limits for binaries varies with the magnitude of the source.  Coupled with the fact than none of the newly detected binaries come from the fainter half of our NICMOS sample, we can consider two hypotheses (assuming that this is a real trend and not a statistical fluke).  First, it is possible that there are fewer binaries for fainter (and presumably smaller) TNOs.  However, no such trend is apparent in the more widely separated binaries found to date (Noll~2003).  Alternatively, it is possible that there are a similar number of binaries in the fainter half of our sample that are undetected because of their combination of faintness and small separation.  If this second explanation is accurate, then the underlying fraction of binaries could be significantly higher than 11\%.  This alternative can be tested with higher resolution, deeper images of a sample of TNOs.

The radial separation of  five of the six new binaries falls between 0.8 and 1.3 pixels.  At a typical geocentric distance of 42 AU, 0.8 pixels (60 milliarcsec) corresponds to 1800 km.  With typical TNOs having radii of 100 km or less, there is ample space for currently undetectable binaries to be hidden.  Indeed, some models of binary formation (e.g. Goldreich et al.~2002)  predict there could be significantly more pairs at smaller angular separations than our current observations can reach.

At the extreme end of smaller separation are contact binaries and bilobate objects.  Such objects, if they exist, are well beyond the capabilities of Earth based imaging systems.  For example, a hypothetical contact binary at 42 AU with two 100 km diameter components would have a maximum angular dimension of 6 mas.  However, it is possible to infer the presence of contact binaries and bilobate objects from their unusual lightcurves.  Sheppard \& Jewitt (2004) have identified one candidate, \QG , from its unusually large-amplitude and long-period lightcurve.  They estimate that 10-15\% of TNOs may be similar systems.  While this interpretation of lightcurve data is not indisputable, it underscores the fact that any binary fraction based on direct imaging is a lower limit.

The NICMOS sample is large enough that we can begin to examine possible differences between the binary fraction in the dynamically cold and excited transneptunian populations.  Table~4 summarizes the statistics for each of the dynamical classes and for the aggregate of the dynamically hot objects.   Six of the nine binaries detected in the NICMOS sample are classical Kuiper Belt objects with inclinations of less than 5 degrees, i.e. the cold classical belt.   The NICMOS sample includes twenty-seven cold classical objects which leads to a computed rate of binaries of 22$\pm {10\atop 6}$\% , clearly higher than the average for the sample.  

When all of the dynamically excited classes, i.e. the scattered disk, resonant objects, hot classicals, and Centaurs are combined we find a binary rate 5.5$\pm {4\atop 2}$\% .  This is a statistically significant factor of 4 lower frequency of binaries compared to the dynamically cold classical disk.  It is interesting to note, however, that all three of the binaries in the dynamically excited population come from the scattered disk objects, one in the near scattered disk and two in the extended scattered disk.   

We propose a possible explanation for the lower fraction of binaries in dynamically hot populations: Objects that have undergone strong scattering events, i.e. the dynamically hot populations, will have lost a higher fraction of primordial binaries than the cold population.  During scattering events binaries will encounter temporarily large impulsive tidal forces that may disrupt weakly bound systems.  A test of this hypothesis is to compare the binary orbit semimajor axes relative to the systems' Hill radii in the two groups.  This requires that the binary orbits be determined; currently only eight orbits are determined (Noll 2003, Margot et al.~2004, Brown et al.~2005), too few to use this test.  

\section{Conclusions}

We have identified six new transneptunian binaries.  These new binaries have been identified by fitting PSFs to a sample of eighty-one unique objects observed with HST that had not already been found to be binary.  Simulations show that with NICMOS we should be able to identify binaries separated by fractions of a pixel using the technique we employed.  By concentrating on a large, homogeneous data set of well-sampled images, we are able to address statistical questions in an unbiased way.  Fifty-four of the objects observed by us with NICMOS are in dynamically excited classes, {\it i.e.}, the scattered disk, hot classicals, resonant objects, and Centaurs.  Of these fifty-four objects only three are binary.  The dynamically cold component of the Kuiper Belt, the classical objects with inclinations below 5 degrees, have a significantly higher fraction of binaries; six of the twenty-seven objects in our unbiased sample are binary.  The factor of 4 difference in binary fractions between cold and excited populations may be due to the selective destruction of binaries during the strong scattering events that are thought to be part of the evolution of excited populations.  By any measure, the fraction of binaries in the transneptunian region appears to be exceptionally high relative to expectations and must be part of any comprehensive model of the formation and evolution of this portion of the solar system.

\acknowledgements {Based on observations made with the NASA/ESA Hubble
Space Telescope. These observations are associated with program 
\#~9386.  Support for program \#~9386 was
provided by NASA through a grant from the Space Telescope Science
Institute, which is operated by the Association of Universities for
Research in Astronomy, Inc., under NASA contract NAS 5-26555.}

\newpage

\noindent{\bf References} 

\ref{Brown, M.~E., Bouchez, A.~H., Rabinowitz, D., Sari, R., Trujillo, C.~A., van Dam, M., Campbell, R., Chin, J., Hartman, S., Johansson, E., Lafon, R., LeMignant, D., Stomski, P., Summers, D., Wizinowich, P.   2005, ApJ Letters, in press.} 

\ref{Christy, J.~W., \& Harrington, R.~S. 1978, AJ, 83, 1005}

\ref {Goldreich, P., Lithwick, Y., \& Sari, R.  2002, Nature, 420, 643}


\ref {Kern, S.~D., \&  Elliot, J.~L. 2005, IAUC, 8526}

\ref {Krist, J.~E., \& Hook, R.  2003, {\it The TinyTim User's Guide}, v.6.1,
Space Telescope Science Institute, Baltimore}

\ref {Margot, J. L., Brown, M. E., Trujillo, C., Sari, R. 2004, DPS 36, 08.03}

\ref {Noll, K.~S.  2003, Earth Moon Planets, 92, 395}

\ref {Noll, K.~S., Luu, J.,\& Gilmore, D.~M. 2000, AJ, 119, 970}

\ref {Noll,ÊK.~S., Stephens,ÊD.~C., Grundy,ÊW.~M., Millis,ÊR.~L., Spencer,ÊJ., Buie,ÊM.~W., Tegler,ÊS.~C., Romanishin,ÊW., \& Cruikshank,ÊD.~P.   2002a,  IAUC, 8034}

\ref {Noll, K., Stephens, D., Grundy, W., Millis, R., Buie, M., Spencer, J.,
Tegler, S., Romanishin, W., \& Cruikshank, D.  2002b, AJ, 124,
3424}

\ref {Noll,ÊK.~S., Stephens,ÊD.ÊC., Cruikshank,ÊD., Grundy,ÊW., Romanishin,ÊW., \& Tegler,ÊS.  2003, IAUC, 8143} 

\ref{Noll, K.~S., Stephens, D.~C., Grundy, W.~M., Osip, D.~J., \& Griffin, I.  2004a, AJ, 128, 2547}

 \ref{Noll, K.~S., Stephens, D.~C., Grundy, W.~M., \& Griffin, I.  2004b, Icarus, 172, 402}

\ref {Sheppard, S.~S., \& Jewitt, D.~C.  2004, AJ, 127, 3023}

\ref {Stephens, D., Noll, K., \& Grundy, W.  2004, IAUC, 8289}

\ref {Veillet, C., Parker, J.~W., Griffin, I., Marsden, B.,
Doressoundiram, A., Buie, M., Tholen, D.~J., Connelley, \& M., Holman,
M.~J.  2002, Nature, 416, 711}

\begin{figure}
\includegraphics[totalheight=0.75\textheight,angle=0]{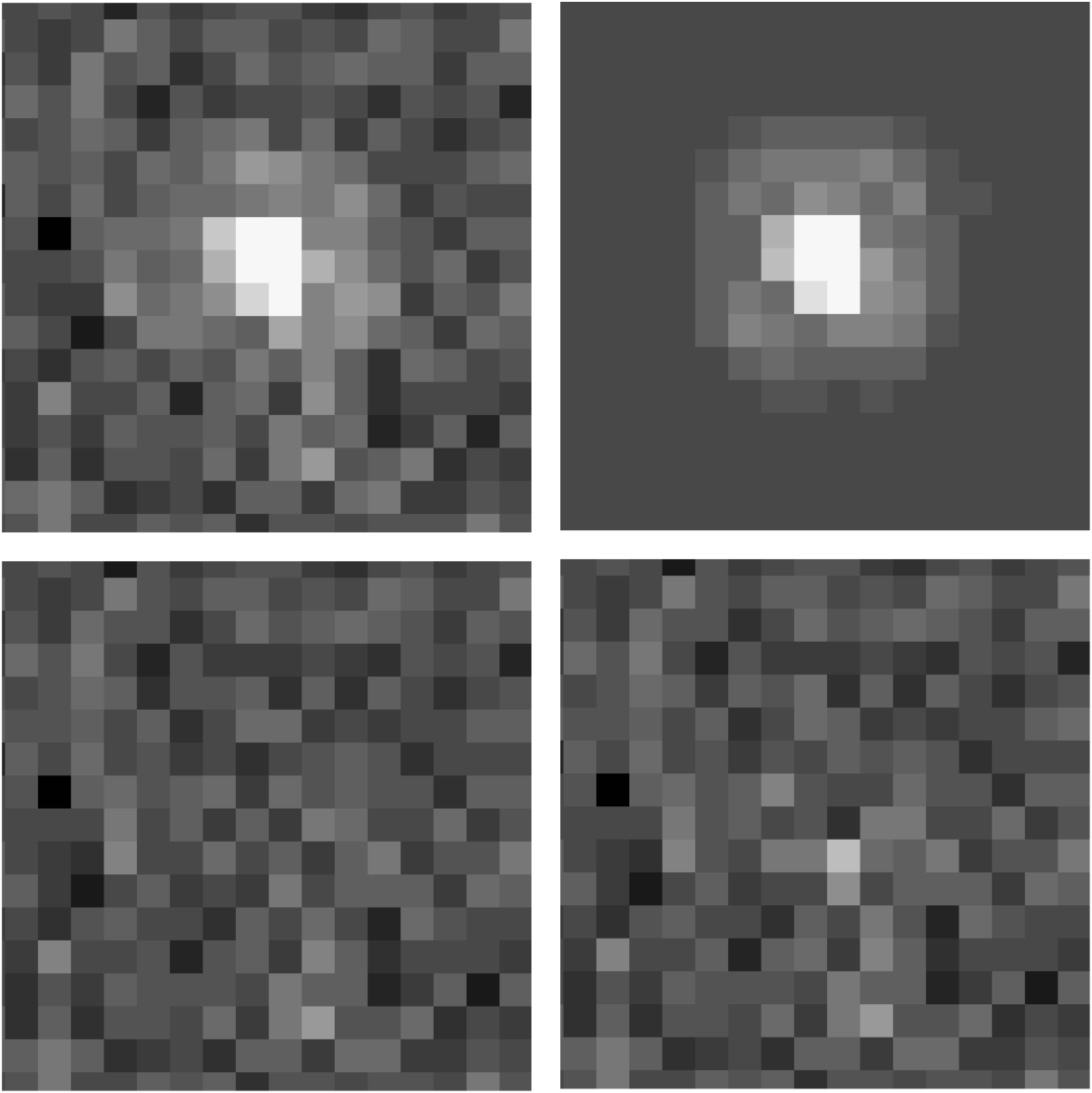}
\caption {A single F160W image of the binary candidate \CM\ is shown in the upper left panel.  The best-fit binary model is shown in the upper right.  The lower left panel shows the result of subtracting the model from the observed data; residuals are indistinguishable from noise.  The residuals left after subtracting the best fitting single PSF from the data are shown in the lower right.  Flux that is significantly above or below the background is apparent in a few pixels. }

\label{fig1}
\end{figure}

\begin{figure}
\includegraphics[totalheight=0.65\textheight,angle=0]{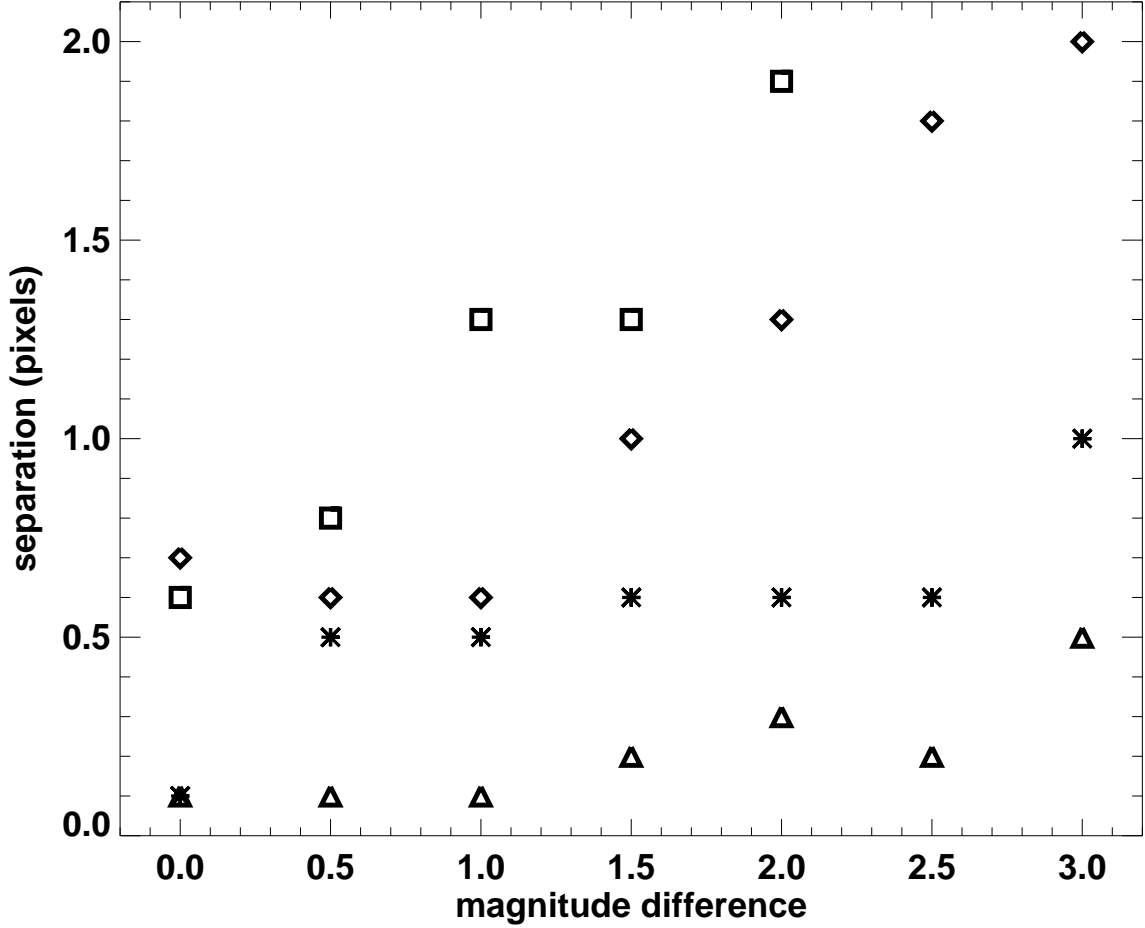}
\caption {Detectability of binaries under various conditions is shown.  Each set of symbols is for a primary of a given H-band magnitude (18.5-triangles; 19.5-asterisks; 20.5-diamonds; 21.5-squares).  The magnitude difference between primary and test secondary is shown on the x-axis.  At each half-magnitude step the separation in pixels at which more than half of the tests resulted in detection of the secondary is plotted for each set.  Noise in each test is appropriate for the NIC2 camera and the integration times used in these observations.}

\label{fig2}
\end{figure}

\begin{figure}
\includegraphics[totalheight=0.5\textheight,angle=0]{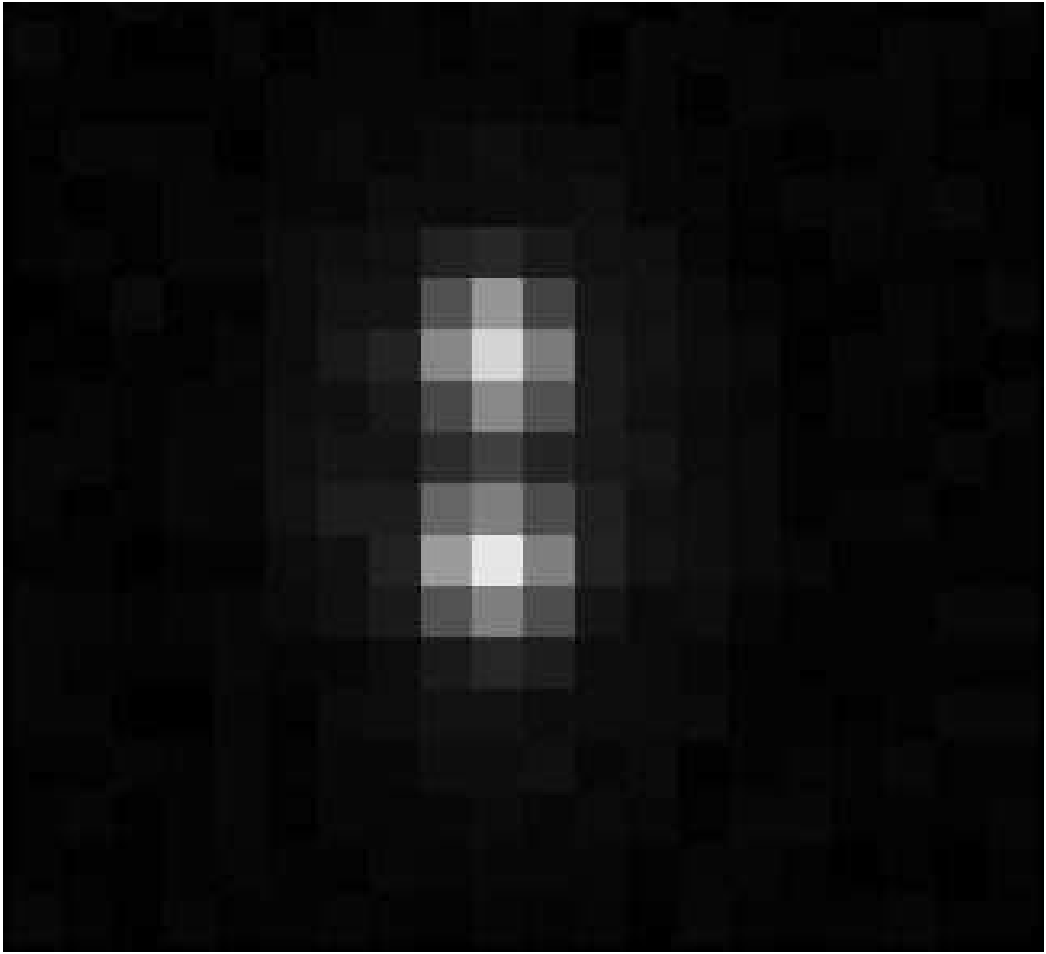}
\caption {Observation of \OJ\ obtained with the Advance Camera High Resolution Camera (ACS HRC) on 24 July 2005.  Each pixel is approximately 25 milliarcsec on a side.  Both components of the binary are clearly resolved with a separation of 100 milliarcsec at this epoch.}

\label{fig3}
\end{figure}

\newpage

\begin{table*}
\label{targetlist}
\begin{tabular}{lcccccl}

\multispan6\hfil{\bf Table 1: Confirmed Binaries in NICMOS Sample}\hfil \cr
\noalign { \vskip 12pt \hrule height 1pt \vskip 1pt \hrule height 1pt \vskip 8pt } 
object & dyn. & date  & \multicolumn{2}{c}{separation}   & $\Delta $mag  & reference  \cr
            & class          & (UT)  &  (arcsec)    & (pixels)        &  (F160W)          &                    \cr
\noalign { \vskip 8pt\hrule height 1pt \vskip 8pt }
\noalign {\bigskip } 
\CF   & C  & 2003 Feb 24.15  & 0.964 $\pm$ 0.004 & 13  & 0.54$\pm$0.09  & this work  \cr 
\noalign {\smallskip}
         &      & 2003 Apr 22.92   & 0.925 $\pm$ 0.004 & 12  & 0.54$\pm$0.07  & this work  \cr 
\noalign {\smallskip}
\CQ  & C  & 2003 May 4.20    & 0.340 $\pm$ 0.005 & 4.5 & 0.58$\pm$0.06 & Noll et al.~(2004a)  \cr 
\noalign {\smallskip}
\RZ   & C  & 2003 Apr 23.14   & 0.192 $\pm$ 0.004 & 2.6 & 0.44$\pm$0.02 & Noll et al.~(2004b)  \cr 
\noalign {\smallskip}
\CQF & C & 2003 Jun 6.02    & 0.183 $\pm$ 0.004 & 2.4 & 0.30$\pm$0.02  & Stephens et al.~(2004) \cr 
\noalign {\smallskip}
\QC  & S  & 2002 Oct 4.66     & 0.130$\pm$ 0.007  & 1.7 & 0.58$\pm$0.03 & Noll et al.~(2002a)  \cr 
\noalign {\smallskip}
         &      & 2003 Apr 22.96   & 0.097$\pm$ 0.009  & 1.3 & 0.69$\pm$0.03 & this work  \cr 
\noalign {\smallskip}
\noalign {\vskip 8pt \hrule height 1pt }  
\noalign {\smallskip}
\multispan7{C = classical, S = scattered}\hfil\cr

\end{tabular} 
\end{table*}

\begin{table*}
\label{targetlist}
\begin{tabular}{lcccccc}
\multispan6\hfil{\bf Table 2: Newly Identified Close Binaries in NICMOS Sample}\hfil \cr
\noalign { \vskip 12pt \hrule height 1pt \vskip 1pt \hrule height 1pt \vskip 8pt } 
object 			& dyn. & date  	& \multicolumn{2}{c}{separation}  	& total mag & $\Delta $mag 	 \cr
            			& class& (UT)  			&  (arcsec)    	& (pix.)  	& (F160W)		&  (F160W)  	 \cr
\noalign { \vskip 8pt\hrule height 1pt \vskip 8pt }
\noalign {\bigskip } 
\CM    				& C	& 2002 Oct 5.20		&  0.088 $\pm$ 0.01  & 1.2   & 21.2	&  1.2      \cr \noalign {\smallskip}
\ph{(12345)} \OJ$^{1,2}$	& C	& 2002  Oct 4.59       	& 0.080 $\pm$ 0.01    & 1.1  & 21.2	& 1.3  	\cr \noalign {\smallskip}
\ph{(12345)} \OJS  		& C	& 2003  Jun 25.72    	&  0.076 $\pm$ 0.01  &  1.0  & 20.7	&  0.8         \cr \noalign {\smallskip}
\CS    				& C	& 2002  Oct 22.98     	&  0.074 $\pm$ 0.01  & 1.0   & 20.0	&  0.1         \cr \noalign {\smallskip}
\YW $^2$   			& S 	& 2002  Oct 25.02     	&  0.061 $\pm$ 0.01   &  0.8 & 19.2	& 1.3    	\cr \noalign {\smallskip}
\TL     				&  S 	& 2002  Nov 9.23       & 0.014 $\pm$ 0.01    & 0.2  & 18.9	& 1.7	 			\cr \noalign {\smallskip}
\noalign { \vskip 12pt \hrule height 1pt \vskip 8pt }  
\noalign {\smallskip}
\multispan7{C = classical, S = scattered}\hfil\cr
\noalign {\smallskip}
\multispan7{$^1$ confirmed with ACS HRC} \hfil\cr
\noalign {\smallskip}
\multispan7{$^2$ two images only}\hfil\cr
\end{tabular} 
\end{table*}
\bigskip

\newpage\begin{table*}
\label{nada}
\begin{tabular}{llll}
\multispan4\hfil{\bf Table 3: NICMOS Targets With No Detectable Binary Companion}\hfil \cr
\noalign { \vskip 12pt \hrule height 1pt \vskip 1pt \hrule height 1pt \vskip 8pt } 
\ph{(12345)} 1993 FW               & (15788) 1993 SB                           & (15789) 1993 SC                        &  (15810) 1994 JR$_1$                \\
(15809) 1994 JS                         & (19255) 1994 VK$_8$                 & (24835) 1995 SM$_{55}$          & (48639) 1995 TL$_8$                 \\        
\ph{(12345)} 1996 KV$_1$      &\ph{(12345)} 1996 RQ$_{20}$	    & (15874) 1996 TL$_{66}$            & (19308) 1996 TO$_{66}$           \\
(15875) 1996 TP$_{66}$          &\ph{(12345)} 1996 TQ$_{66}$     & (20161) 1996 TR$_{66}$           &\ph{(12345)} 1996 TS$_{66}$    \\
(15883) 1997 CR$_{29}$         &\ph{(12345)} 1997 CT$_{29}$     & (33001) 1997 CU$_{29}$          &\ph{(12345)} 1997 CV$_{29}$    \\
\ph{(12345)} 1997 RT$_{5}$    &\ph{(12345)} 1998 FS$_{144}$   &\ph{(12345)} 1998 KG$_{62}$   &\ph{(12345)} 1998 KY$_{61}$    \\
\ph{(12345)} 1998 UU$_{43}$ & (33340) 1998 VG$_{44}$            &  (19521) 1998 WH$_{24}$        &\ph{(12345)} 1998 WX$_{31}$   \\
\ph{(12345)} 1998 WY$_{24}$ &\ph{(12345)}  1998 WZ$_{31}$   &\ph{(12345)} 1999 CD$_{158}$ &\ph{(12345)} 1999 CF$_{119}$ \\
\ph{(12345)} 1999 CF$_{119}$&\ph{(12345)}   1999 CH$_{119}$&\ph{(12345)} 1999 CJ$_{119}$ &\ph{(12345)} 1999 CL$_{119}$ \\
\ph{(12345)} 1999 CX$_{131}$& (26375) 1999 DE$_9$                &\ph{(12345)} 1999 HC$_{12}$   &\ph{(12345)} 1999 HJ$_{12}$    \\
\ph{(12345)} 1999 OD$_4$       &\ph{(12345)}   1999 OE$_4$       &  (66452) 1999 OF$_4$               &  (86047) 1999 OY$_3$              \\
\ph{(12345)} 1999 RC$_{215}$&  (86177) 1999 RY$_{215}$        &\ph{(12345)} 2000 AF$_{255}$ &\ph{(12345)} 2000 CE$_{105}$ \\
\ph{(12345)} 2000 CG$_{105}$&\ph{(12345)} 2000 CK$_{105}$  &\ph{(12345)} 2000 CL$_{104}$ &\ph{(12345)} 2000 CO$_{105}$\\ 
\ph{(12345)} 2000 CP$_{104}$&\ph{(12345)} 2000 CQ$_{105}$  &\ph{(12345)} 2000 CR$_{105}$&  (60621) 2000 FE$_8$               \\
\ph{(12345)} 2000 OK$_{67}$  &\ph{(12345)} 2000 OU$_{69}$    &\ph{(12345)} 2000 PD$_{30}$   &\ph{(12345)} 2000 PE$_{30}$    \\
\ph{(12345)} 2000 PH$_{30}$  &  (54520) 2000 PJ$_{30}$            & (45802) 2000 PV$_{29}$          &\ph{(12345)} 2001 KD$_{77}$    \\
\ph{(12345)} 2001 KY$_{76}$  &\ph{(12345)} 2001 OG$_{109}$  &\ph{(12345)} 2001 OK$_{108}$&\ph{(12345)} 2001 QX$_{322}$  \\
\ph{(12345)} 2001 UR$_{163}$&\ph{(12345)} 2001 XS$_{254}$  &\ph{(12345)} 2001 XU$_{254}$                                                             \\
\noalign { \vskip 12pt \hrule height 1pt \vskip 8pt } 
\end{tabular} 
\end{table*}

\newpage\begin{table*}
\label{frequency}
\begin{tabular}{lcccccc}
\multispan7\hfil{\bf Table 4: Fraction of Binaries in Dynamically Hot and Cold Populations}\hfil \cr
\noalign { \vskip 12pt \hrule height 1pt \vskip 1pt \hrule height 1pt \vskip 8pt } 
 & {\bf cold}	               & {\bf hot}    & \multicolumn{4}{c}{hot}                                                                                \\
 & classical i$<$5    &  all        & classical i$>$5    &    resonant   & scattered  & centaurs  \\        
 & {\underbar { \ph{ classicalxx }}} & {\underbar{\ph{  all  }}} & \multicolumn{4}{c}{\underbar{\phantom{      classical      resonant        scattered          centaurs   morespace  }}} \\ 
NICMOS	& 27  		& 54  	 	& 10			& 17			& 26			&   1     \\
binaries    	& 6			&   3			&  0   		& 0      		& 3			&   0     \\
fraction$^*$ (\%)& {\bf 22$\pm {10\atop 5}$} & {\bf 5.5$\pm {4\atop 2}$} 	& -	& -	& 11.5$\pm {9\atop 4}$  	&   -	    \\
\noalign { \vskip 12pt \hrule height 1pt \vskip 8pt } 
\multispan5{$^*$ All objects combined yield a binary fraction of 11$\pm {5\atop 2}$\% } \hfil\cr
\end{tabular} 
\end{table*}

\end{document}